\documentclass[11pt,a4paper]{article}
\usepackage{epsfig}

\newlength{\dinwidth}
\newlength{\dinmargin}
\def\eq#1{{Eq.~(\ref{#1})}}
\newcommand{\Le}{\left(}
\newcommand{\Ra}{\right)}

\newcommand{\beq}{\begin{equation}}
\newcommand{\eeq}{\end{equation}}
\newcommand{\beqar}[1]{\begin{eqnarray}\label{#1}}
\newcommand{\eeqar}{\end{eqnarray}}
\begin{document}

\title {{~}\\
{\Large \bf On asymptotic solutions of RFT in zero transverse dimensions }\\}
\author{
{~}\\
{~}\\
{\large 
S.~Bondarenko$\,{}^{a)}\,$\thanks{Email: sergeyb@ariel.ac.il}, 
\hspace{1ex}
L.~Horwitz$\,{}^{a),b)}\,$\thanks{E-mail: larry@post.tau.ac.il},
\hspace{1ex}
J.~Levitan$\,{}^{a)}\,$\thanks{E-mail: levitan@ariel.ac.il}
\hspace{1ex}
A.~Yahalom$\,{}^{a)}\,$\thanks{Email: asher@fluidex-cfd.com}
}
\\[10mm]
{\it\normalsize ${}^{a)}$ Ariel University Center}\\
{\it\normalsize $^{b)}$ Tel Aviv University}\\}

\maketitle
\thispagestyle{empty}

\begin{abstract}
An investigation of dynamical properties of solutions of toy model of interacting Pomerons with  triple
vertex in zero transverse dimension is performed. 
Stable points  and corresponding solutions at the limit of large rapidity are studied in the framework of given model.
A presence of closed cycles in solutions is
discussed as well as
an application of obtained results for the case of interacting QCD Pomerons.
\end{abstract}

\newpage

\section{Introduction}

A calculating of amplitudes of QCD high-energy scattering processes
is a complex task based on different methods and approaches
of high-energy physics, see
\cite{bfkl,bfklsum,nlbfkl,vert1,jimwalk}.
In general this task is very complicated, and, therefore, the simplified models which 
have dynamics similar to QCD ones could serve as a good polygon for the different ideas and methods check.  
This is the reason why RFT-0 (Reggeon Field Theory
in zero transverse dimensions) model is attracted much interest during few last years,
see   \cite{0dimsol,BondRF}. The approach, which was formulated 
and studied a long time ago, even before the QCD era ( see \cite{0dimc}), has a dynamics which is very similar to the 
solutions obtained in framework of interacting QCD Pomerons, see \cite{Braun,BonMot}.

In the paper \cite{BonMot} was shown, that we could understand the real RFT dynamics solving  equations of motion of
RFT-0. Quantum amplitudes in  RFT-0 model framework are also calculable, see \cite{BondRF}, and this is the only source 
for understanding, if only partially, of quantum dynamics of RFT based on QCD Pomerons. Nevertheless, even at classical level,
there are some open questions which are interesting from point of view of QCD high-energy description of the process. 
These questions are about the asymptotic solutions of equations of motion, their stability and properties, see more details in \cite{BonMot}. In contrast to what we
have in RFT in two transverse dimensions, the classical solutions of RFT-0 could be investigated analytically and achieved results 
could be applicable also in high-energy QCD.

The paper is organized as follows. In the next section we consider a way to approach asymptotic solutions of the model whereas in the Section 3 
we analyze steady states of equations of motion. In Section 4 we discuss the interconnections between the steady-states and asymptotic solutions,
relating the stable points of the equations with corresponding asymptotic solutions. Section 5 is dedicated to additional analysis of a stability
of some asymptotic solutions. In Section 6 we investigate a possibility of periodic limited cycles of the solutions and Section 7 is a 
discussion of results and conclusion of the paper.

\section{Asymptotic behavior of equation of motion}

The Hamiltonian of our problem and corresponding equation of motions have the following form:
\beq\label{InHam}
H=\mu\,q\,p\,-\,\gamma\,p^2\,q\,-\,\gamma\,p\,q^2\,
\eeq
and
\beq
\dot{q}\,=\mu\,q\,-\,\gamma\,q^2\,-\,2\,\gamma\,p\,q\,
\eeq
\beq
\dot{p}\,=\,-\mu\,p\,+\,\gamma\,p^2\,+\,2\,\gamma\,p\,q\,,
\eeq
see \cite{BondRF,0dimc}, where $\mu$ is an intercept of bare Pomeron and $\lambda$ is a vertex of triple Pomeron interactions.
The "time" variable of the equations is a rapidity , i.e. $\dot{q}=\frac{dq}{dy}$
with $y\,=\,\ln\,(s/s_{0})\,$ and with $s$ as a squared total energy of the process..
The boundary conditions are symmetric
\beq\label{InCon}
q(y=0)=A\,,\,\,\,p(y=Y)=A\,\,\,
\eeq
where $Y$ is the end of the "time" interval, i.e. Y is a total rapidity of the scattering process. We also request,
that an amplitude of the "scattering" process, which depends on $q$ and $p$ variables, will preserve the so called "target-projectile" symmetry
\beq\label{SymCon}
Ampl.(q(y,A),p(y,A))=Ampl.(q(Y-y,A),p(Y-y,A))\,
\eeq
see in \cite{BonMot} the definition of the scattering amplitude in the RFT.

In order to qualitatively understand behavior of the equations at large rapidities
we rescale the rapidity and vertex of the problem:
\beq
y\rightarrow\,\mu\,y\,,\,\,\,\lambda\,=\,\gamma\,/\,\mu
\eeq
obtaining new equation of motion
\beq\label{Motion1}
\dot{q}\,=\,q\,-\,\lambda\,q^2\,-\,2\,\lambda\,p\,q\,
\eeq
\beq\label{Motion2}
\dot{p}\,=\,-\,p\,+\,\lambda\,\,p^2\,+\,2\,\lambda\,p\,q\,.
\eeq
We see, that in the rescaled equation the asymptotic limit for the rescaled total rapidity $Y$  
$$
\,Y_{Rescaled}\,=\,\mu\,Y\,\rightarrow\,\infty
$$
could be achieved also by the limit
$$
\mu\,\rightarrow\,\infty\,,\,\,\,\,\,\gamma\,\propto\,\alpha_{s}^2\,\rightarrow\,\infty
$$
where also
$$
\lambda\,=\,\gamma\,/\,\mu\,\propto\,\alpha_{s}\,\rightarrow\,\infty\,.
$$
It means, that asymptotic solution of equations of motion for the case $Y\,\rightarrow\,\infty\,\,$
are the same as solutions in the limit $\lambda\rightarrow\,\infty\,$. Moreover, because our function must be analytical function of $\alpha_s$,
the solutions will be valid also when $Y\,\rightarrow\,\infty\,\,$ at finite $\lambda$. We note also, that 
due the initial condition ~\eq{InCon}, 
the asymptotic behavior of the $q\,=\,q(y,A)$ 
is achieved in the limit $y\,\rightarrow\,Y\,\rightarrow\,\infty$ whereas asymptotic of $p\,=\,p(y,A)$ function is lies
in the limit $y\,\rightarrow\,0$ with $\,Y\,\rightarrow\,\infty$.

The rescaled Hamiltonian of the problem now has the following form:
\beq\label{En}
H\,=\,q\,p\,-\,\lambda\,p^2\,q\,-\,\lambda\,p\,q^2\,=\,E
\eeq 
with equation of motion
\beq\label{Motion11}
\dot{q}\,=\,q\,-\,\lambda\,q^2\,-\,2\,\lambda\,p\,q\,
\eeq
\beq\label{Motion22}
\dot{p}\,=\,-\,p\,+\,\lambda\,p^2\,+\,2\,\lambda\,p\,q\, 
\eeq
where 
$$
\,y\,\rightarrow\,\mu\,y\,\,\,\,\,\,\,\,\,E\,\rightarrow\,E\,/\,\mu\,
$$
in equations of motion and Hamiltonian \footnote{Further, we shell use everywhere the sign E instead $E\,/\,\mu\,$.}.
The solution of the equation ~\eq{En} for variables $q$ and $p$ at the limit of large $\lambda$ are
two pairs of the functions:
\beq\label{AsSolQ}
q_{1,2}\,=\,\Le\,\frac{1}{\lambda}\,-\,p\,-\,\frac{E}{\lambda\,p^2}\,,\,\,\frac{E}{\lambda\,p^2}\,\Ra
\eeq
and
\beq\label{AsSolP}
p_{1,2}\,=\,\Le\,\frac{1}{\lambda}\,-\,q\,-\,\frac{E}{\lambda\,q^2}\,,\,\,\frac{E}{\lambda\,q^2}\,\Ra\,.
\eeq
Further, as any pair of asymptotic solution $\Le\,q\,,\,p\,\Ra_{i}$  of equation motion ~\eq{Motion11}\,\,-~\eq{Motion22}, 
we  will consider the
pairs $\Le\,q_{i}\,,\,p_{i}\,\Ra\,\,\,\,i\,=\,1..2\,$ which contain functions ~\eq{AsSolQ}\,\,-~\eq{AsSolP}.

\section{Steady states of equations of motion}

The definition of steady states of the equation of motion is a standard:
\beq
\,q\,\Le\,1\,-\,\lambda\,q\,-\,2\,\lambda\,p\,\Ra\,=\,0
\eeq
\beq
\,-\,p\,\Le\,1\,-\,\lambda\,p\,+\,2\,\lambda\,q\,\Ra\,=\,0
\eeq
Due the fact of absence of rapidity dependence in these equations we assume that at least part of solutions (steady states)
of these equations describe
the solutions of equation of motion in the asymptotic limit $Y\,\rightarrow\,\infty$, whereas 
$y\,\rightarrow\,Y$ for the $q\,=\,q(y,A)$ and $y\,\rightarrow\,0$ for the $p\,=\,p(y,A)$

The non-trivial steady states of the equations could be easily found
\beq\label{StabP1}
\Le\,q_{St}\,,p_{St}\,\Ra_{1}\,=\,\Le\,\frac{1}{\lambda}\,,\,0\,\Ra\,
\eeq
\beq\label{StabP2}
\Le\,q_{St}\,,p_{St}\,\Ra_{2}\,=\,\Le\,0\,,\,\frac{1}{\lambda}\,\Ra\,
\eeq
\beq\label{StabP3}
\Le\,q_{St}\,,p_{St}\,\Ra_{3}\,=\,\Le\,\frac{1}{3\,\lambda}\,,\,\frac{1}{3\,\lambda}\,\Ra\,
\eeq
In general, the linearized analysis of the equations of motion around these states could be performed.
The Jacobian matrix 
\beq
J\,=\,\Le
\begin{array}{cc}
1\,-\,2\,\lambda\,q\,-\,2\,\lambda\,p & -2\,\lambda\,q \\
2\,\lambda\,p & -\,1\,+\,2\,\lambda\,q\,+\,2\,\lambda\,p \\
\end{array}\Ra
\eeq
lead to the following eigenvalues for these three steady states
\beq
\begin{array}{ccc}
\lambda_{1}^{12}\,=\,\pm\,1\,\,\, & \,\,\,\lambda_{2}^{12}\,=\,\pm\,1\,\,\, & \,\,\,\lambda_{3}^{12}\,=\,\pm\,\frac{\imath}{\sqrt{3}}\,. \\
\nonumber\end{array}
\eeq 
Unfortunately, the fact that, for example, the first two steady states are saddles does not say a lot about particular
corresponding asymptotic solutions which also must satisfy boundary conditions ~\eq{InCon}. 
Therefore, in the next two section we will study these steady states in correspondence with the solutions 
~\eq{AsSolQ}\,\,-~\eq{AsSolP} in the sense of convergence of asymptotic solutions to the corresponded steady state in the 
limit of asymptotically large  rapidity. In the case of the convergence existing we could call a steady state as a stable point.

\section{The $\Le\,q_{St}\,,p_{St}\,\Ra_{1}\,$ and $\Le\,q_{St}\,,p_{St}\,\Ra_{2}\,$ 
steady states and corresponding solutions of equation of motion}

In this section we compare the solutions ~\eq{AsSolQ}\,\,-~\eq{AsSolP} with the results of the steady states analysis
~\eq{StabP1}\,\,-~\eq{StabP2}. We note, that
the energy, which corresponds to these steady states is zero:
\beq
E\,=\Le\,p\,q\,-\,\lambda\,p\,q^2\,\Ra_{\{\Le\,q_{St}\,,p_{St}\,\Ra_{1}\,,\,\Le\,q_{St}\,,p_{St}\,\Ra_{2}\,\}}\,=\,0\,.
\eeq
Therefore, in the limit $E\,\rightarrow\,0$ a correspondence
between these states and the asymptotic solution is established easily:
\beq\label{Sol1}
\Le\,q_{St}\,,p_{St}\,\Ra_{1}\,=\,\Le\,\frac{1}{\lambda}\,,\,0\,\Ra\,=\,\Le\,q_{1}\,,\,p_{2}\,\Ra_{E=0}
\eeq
and correspondingly
\beq\label{Sol2}
\Le\,q_{St}\,,p_{St}\,\Ra_{2}\,=\,\Le\,0\,,\,\frac{1}{\lambda}\,\Ra\,=\,\Le\,q_{2}\,,\,p_{1}\,\Ra_{E=0}\,.
\eeq
In order to find the asymptotic solution which corresponds to the steady states/stable points of ~\eq{Sol1}
(we will look for the solution for the states of ~\eq{Sol1}, the solution
for ~\eq{Sol2} could be found similarly) we will consider the equation ~\eq{Motion1} as the equation 
consisting of two parts. The first part provides a fully integrable equation, whereas the second part, $\,2\,\lambda\,q\,p\,$, 
is a small perturbation around $\,p\,\propto\,0$. Keeping only leading terms in both ~\eq{Motion1}\,\,-~\eq{Motion2}
we arrive, therefore, to the following system of equations:
\beq\label{ShortEq1}
\dot{q}\,=\,q\,-\,\lambda\,q^2\,
\eeq
\beq\label{ShortEq2}
\dot{p}\,=\,-\,p\,+\,2\,\lambda\,p\,q\,.
\eeq
The consistency of the approximation we check thereafter
by taking asymptotic limits in the pair $(q,p)$ of solution ans comparison of obtained result with conditions 
~\eq{Sol1} or ~\eq{Sol2}.

Integration of the system ~\eq{ShortEq1}\,-\,~\eq{ShortEq2} with given initial conditions ~\eq{InCon} leads to the following functions:
\beq\label{FullSolQ1}
q(y,A)\,=\,\frac{A\,e^{y}}{A\,\lambda\,\Le\,e^{y}\,-\,1\Ra\,+\,1}
\eeq
\beq\label{FullSolP1}
p(y,A)\,=\,A\,e^{Y-y}\,\Le\,\frac{A\,\lambda\,\Le\,e^{y}\,-\,1\Ra\,+\,1}{A\,\lambda\,\Le\,e^{Y}\,-\,1\Ra\,+\,1}\Ra^{2}\,.
\eeq
In order to find the asymptotic behavior of the solutions we take the limit $Y\,\rightarrow\,\infty$ and finite $\lambda$.
We see, that in this limit, where $y\,\rightarrow\,Y\,\rightarrow\,\infty$ for $q$ and $y\,\rightarrow\,0\,,\,Y\rightarrow\,\infty$ for $p$,
we obtain
\beq\label{SolQ1}
q(y,A)\,=\,\frac{A\,e^{y}}{A\,\lambda\,\Le\,e^{y}\,-\,1\Ra\,+\,1}\,\stackrel{y\rightarrow\,Y\rightarrow\infty}{\longrightarrow}\,\frac{1}{\lambda}
\eeq
\beq\label{SolP1}
p(y,A)\,=\,A\,e^{Y-y}\,\Le\,\frac{A\,\lambda\,\Le\,e^{y}\,-\,1\Ra\,+\,1}{A\,\lambda\,\Le\,e^{Y}\,-\,1\Ra\,+\,1}\Ra^{2}\,\stackrel{y\rightarrow\,0\,,Y\rightarrow\infty}{\longrightarrow}\,\frac{\,e^{-Y}}{A\,\lambda^{2}}\,\longrightarrow\,0\,.
\eeq
These asymptotic values coincide with the asymptotic values $\,\Le\,q_{1}\,,\,p_{2}\,\Ra$ from  
~\eq{AsSolQ}\,-~\eq{AsSolP} and stable points ~\eq{StabP1}. As expected, taking firstly 
the limit $\lambda\,\rightarrow\,\infty$, we will come to the same stable point ~\eq{AsSolQ}\,-~\eq{AsSolP}:
\beq\label{SolQ2}
q(y,A)\,=\,\frac{A\,e^{y}}{A\,\lambda\,\Le\,e^{y}\,-\,1\Ra\,+\,1}\,
\,\stackrel{\lambda\rightarrow\infty\,,y\rightarrow\,Y\rightarrow\,\infty}{\longrightarrow}\,
\frac{1}{\lambda}
\eeq
\beq\label{SolP2}
p(y,A)\,=\,A\,e^{Y-y}\,\Le\,\frac{A\,\lambda\,\Le\,e^{y}\,-\,1\Ra\,+\,1}{A\,\lambda\,\Le\,e^{Y}\,-\,1\Ra\,+\,1}\Ra^{2}\,\stackrel{\lambda\rightarrow\infty\,,y\rightarrow\,0\,Y\rightarrow\,\infty}{\longrightarrow}\,A\,e^{-Y}\,\rightarrow\,0
\eeq 
in the limit $Y_{Rescaled}\,\propto\,\alpha_{s}\,Y\rightarrow\,\infty$.

The second solution, which corresponds to the stable point $\Le\,q_{St}\,,p_{St}\,\Ra_{2}\,$ could be found similarly.
It has the following form
\beq\label{FullSolQ2}
q(y,A)\,=\,\,A\,e^{y}\,\Le\,\frac{A\,\lambda\,\Le\,e^{Y-y}\,-\,1\Ra\,+\,1}{A\,\lambda\,\Le\,e^{Y}\,-\,1\Ra\,+\,1}\Ra^{2}
\eeq
\beq\label{FullSolP2}
p(y,A)\,=\,\frac{A\,e^{Y-y}}{A\,\lambda\,\Le\,e^{Y-y}\,-\,1\Ra\,+\,1}\,.
\eeq
We also see, that an amplitude consisting of the pairs of solutions, ~\eq{FullSolQ1}-~\eq{FullSolP1} and ~\eq{FullSolQ2}-~\eq{FullSolP2}, will preserve the symmetry of the problem ~\eq{SymCon}

\section{The $\Le\,q_{St}\,,p_{St}\,\Ra_{3}\,$ 
steady state and corresponding solution of equation of motion}

We search an asymptotic solution which characterized by following steady state:
\beq\label{ThirdSP}
\Le\,q_{St}\,,p_{St}\,\Ra_{3}\,=\,\Le\,\frac{1}{3\,\lambda}\,,\,\frac{1}{3\,\lambda}\,\Ra\,.
\eeq
As a  pair of asymptotic solutions $(q\,,\,p)$, which could corresponds to this state, we take the following combination
\beq\label{AsSol3}
\Le\,q_{1}\,,\,p_{1}\,\Ra\,=\,\Le\,\frac{1}{\lambda}\,-\,p\,-\,\frac{E}{\lambda\,p^2}\,,\frac{1}{\lambda}\,-\,q\,-\,\frac{E}{\lambda\,q^2}\,\Ra\,,
\eeq
and our further task is a check of the self-consistency of this assumption. 
The Hamiltonian for this steady state/stable point is not zero:
\beq
\,H(\,q=1/3\lambda\,,\,p=1/3\lambda\,)\,=\,E\,=\,\frac{1}{27\lambda^2}\,,
\eeq
so we need to find solutions of equations of motion which provide this non-zero energy.

Let's consider full equation for the $q$ variable
and, as we did previously above, we will expand the last "perturbative" term in the equation taking there $p\,=\,p_{1}$ from ~\eq{AsSol3} 
\beq
\dot{q}\,=\,q\,-\,\lambda\,q^2\,-\,2\,\lambda\,p\,q\,=\,\,q\,-\,\lambda\,q^2\,-\,2\,\lambda\,p_{1}\,q\,
\eeq
We obtain
\beq\label{Sol3}
\dot{q}\,=\,q\,-\,\lambda\,q^2\,-\,2\,\lambda\,p_{1}\,q\,=\,\,q\,-\,\lambda\,q^2\,-\,2\,\lambda\,q\,\Le\,
\frac{1}{\lambda}\,-\,q\,-\,\frac{E}{\lambda\,q^2}\,\Ra\,=\,-\,q\,+\,\lambda\,q^2\,+\,\frac{2\,E}{\,q}\,.
\eeq
Again, considering the last term as an perturbation, we take there $q=q_{St}$ from ~\eq{ThirdSP} and rescaling rapidity $\mu\,y\,\rightarrow\,\mu\,\lambda\,y$ we get finally:
\beq\label{Sol4}
\dot{q}\,=\,q^2\,-\,q\,/\,\lambda\,+\,6\,E\,.
\eeq
The solution of this equation with the initial condition given by ~\eq{InCon} is
\beq\label{SolQ3}
q=\frac{1}{2\lambda}\,+\,\acute{E}\,\frac{A\,-\,1/(2\lambda)\,+\,\acute{E}\,+\,e^{2\acute{E}\,y}\,\Le\,A\,-\,1/(2\lambda)\,-\,\acute{E}\,\Ra}
{A\,-\,1/(2\lambda)\,+\,\acute{E}\,-\,e^{2\acute{E}\,y}\,\Le\,A\,-\,1/(2\lambda)\,-\,\acute{E}\,\Ra}
\eeq
where
\beq\label{EnCon1}
\acute{E}\,=\,\sqrt{\frac{1}{4\lambda^2}-6E}\,
\eeq
and rapidity is $\gamma\,y\,$.
In the asymptotic limit $y\,\rightarrow\,Y\,\rightarrow\,\infty$ we obtain a simple expression
\beq\label{AsQ}
q=\frac{1}{2\lambda}\,-\,\acute{E}\,.
\eeq
For the $p$ variable we similarly obtain
\beq\label{SolP3}
p=\frac{1}{2\lambda}\,+\,\acute{E}\,\frac{A\,-\,1/(2\lambda)\,+\,\acute{E}\,+\,e^{2\acute{E}\,(Y-y)}\,\Le\,A\,-\,1/(2\lambda)\,-\,\acute{E}\,\Ra}
{A\,-\,1/(2\lambda)\,+\,\acute{E}\,-\,e^{2\acute{E}\,(Y-y)}\,\Le\,A\,-\,1/(2\lambda)\,-\,\acute{E}\,\Ra}
\eeq
that in the asymptotic limit of variable $p$, which is $y\,\rightarrow\,0\,,\,\,Y\,\rightarrow\,\infty$, gives
\beq\label{AsP}
p=\frac{1}{2\lambda}\,-\,\acute{E}\,.
\eeq
How, for the verification of the self-consistency of the solution, we back to the asymptotic pair of solutions ~\eq{AsSol3}.
We have
\beq
q\,=\,\frac{1}{2\lambda}\,-\,\acute{E}\,=\,\frac{1}{\lambda}\,-\,p\,-\,\frac{E}{\lambda\,p^2}\,.
\eeq
Taking value of $p$ from ~\eq{AsP} and value of $E$ from ~\eq{EnCon1} we obtain
\beq
-2\,\lambda\,\acute{E}\,\Le\,1/(2\lambda)\,-\,\acute{E}\,\Ra^{2}\,=\,\,\acute{E}^{2}/6\,-\,1/(24\,\lambda^2)\,.
\eeq
Among three solutions of this equation there is a solution
\beq
\acute{E}\,=\,\frac{1}{6\lambda}
\eeq
which gives
\beq
(q_{1}\,,\,p_{1}\,)\,=\,(\frac{1}{2\lambda}\,-\,\acute{E}\,,\,\frac{1}{2\lambda}\,-\,\acute{E})_{\acute{E}=\frac{1}{6\lambda}}\,=\,
\,\Le\,\frac{1}{3\,\lambda}\,,\,\frac{1}{3\,\lambda}\,\Ra\,=\Le\,q_{St}\,,p_{St}\,\Ra_{3}\,
\eeq
and
\beq
E\,=\,1/(24\,\lambda^2)\,-\,\acute{E}^{2}/6\,=\,\frac{1}{27\lambda^2}\,=\,H(\,q=1/3\lambda\,,\,p=1/3\lambda\,)\,.
\eeq
Clearly, the required amplitude's  symmetry ~\eq{SymCon} for the ~\eq{SolQ3}, ~\eq{SolP3}  functions is also preserved.

 Thereby we proved that our solution is self-consistent ant that 
solution ~\eq{AsSol3} (i.e. functions ~\eq{SolQ3},~\eq{SolP3}) is related with the ~\eq{StabP3} stable point.
Other pairs of asymptotic solutions for $E\,\neq\,0$, such as $\Le\,q_{2}\,,\,p_{2}\,\Ra\,$, $\Le\,q_{1}\,,\,p_{2}\,\Ra\,$
and $\Le\,q_{2}\,,\,p_{1}\,\Ra\,$, do not satisfy the condition of correspondence of their large rapidity limits with the steady state 
~\eq{StabP3}.

\section{Stability analysis and Lyapunov function}
\,\,\,

Obtained asymptotic solutions are stable in the sense of their behavior at asymptotically large rapidity.
Nevertheless, it is interesting to check the asymptotic stability of the solution from the point view 
of Lyapunov function, namely, we will try to construct the local Lyapunov functions for each 
stable point and corresponding asymptotic solution obtaining the stability criteria from this side of the problem.

We will look for the local Lyapunov functions, it means that we will search for the functions which have following
properties:
\begin{enumerate}
\item 
first property:
\beq
F(q,p)\,\geq\,0
\eeq	
for the $q$ and $p$ defined in neighborhood region of stable point;	
\item
second property:
\beq
F(q,p)_{q=q_{SP},p=p_{SP}}\,=\,0\,;
\eeq
\item
third property:
\beq
\Le\,\frac{dF(q,p)}{dy}\,\Ra\,<\,0
\eeq
in neighborhood region of stable point.
\end{enumerate}

\subsection{Lyapunov function for ~\eq{StabP1}-~\eq{StabP2} stable points}

We consider stable points ~\eq{StabP1}
\beq\label{StabP11}
\Le\,q_{St}\,,p_{St}\,\Ra_{1}\,=\,\Le\,\frac{1}{\lambda}\,,\,0\,\Ra\,
\eeq
and corresponding asymptotic solution $(q_{1},p_{2})$ defined by ~\eq{FullSolQ1}-\,~\eq{FullSolP1}.
For the stable point $\Le\,q_{St}\,,p_{St}\,\Ra_{1}\,$ we consider 
a following  function as a candidate
for the Lyapunov function:
\beq\label{Lyap1}
F(q,p)\,=\,H\,-\,p^n\,+\,\lambda\,p^2\,q
\eeq
where $H$ is the Hamiltonian ~\eq{En} and $n\,>\,3$. Now we will check properties of this function as a Lypunov one.
\begin{enumerate}
\item 
We look for the behavior of the Lyapunov function around the stable point and we have at leading order at
asymptotically large $Y$ :
\beq
F(q,p)_{(q_{1},p_{2})}\,\,=\Le\,q\,p\,-\,\,\lambda\,p\,q^2\,-\,p^n\,\Ra_{(q_{1},p_{2})}\,=\,
\frac{e^{-2\,Y}\,\Le\,1\,-\,\lambda\,A\,\Ra}{A^2\,\lambda^4}\,>\,0\,,
\eeq	
the condition $n>3$ is used here for the derivation of the answer and where we assume $\,\lambda\,A\,<\,1\,.$
\item
Checking second request for the Lyapunov function we obtain
\beq
F(q=\frac{1}{\lambda},p=0)\,=\,0\,
\eeq
simply by definition of the function which is proportional to $p$.
\item
Third property of our function is the following
\beq
\Le\,\frac{dF(q,p)}{dy}\,\Ra_{(q_{1},p_{2})}\,=\Le\,q\,p\,-\,\,\lambda\,p\,q^2\,-\,p^n\,\Ra^{'}_{(q_{1},p_{2})}\,=\,\Le\,
\dot{H_{0}}\,-\,n\,p^{n-1}\,\dot{p}\,\,\Ra_{(q_{1},p_{2})}\,.
\eeq
Here $H_{0}\,=\,q\,p\,-\,\,\lambda\,p\,q^2\,$ is a "fan" Hamiltonian with precise solution 
of the corresponding equations of motion in the form functions 
~\eq{FullSolQ1}-\,~\eq{FullSolP1}. Therefore, we have $\dot{H_{0}}\,=\,0$  and obtain:
\beq
\Le\,\frac{dF(q,p)}{dy}\,\Ra_{(q_{1},p_{2})}\,=\,-\Le\,n\,p^{n-1}\,\dot{p}\,\,\Ra_{(q_{1},p_{2})}=\,-n\,p^{n}\,\Le\,-\,1\,+\,\lambda\,p\,+\,
2\,\lambda\,q\,\Ra_{(q_{1},p_{2})}\,.
\eeq
In leading order, around $(q_{1}\,=\,\frac{1}{\lambda},p_{2}\,=\,\epsilon)$, we obtain:
\beq
\Le\,\frac{dF(q,p)}{dy}\,\Ra_{(q_{1},p_{2})}\,=\,-\,n\,\epsilon^{n}\,=\,-n\,\frac{e^{-n\,Y}}{A^{n}\,\lambda^{2n}}\,<\,0\,.
\eeq
\end{enumerate}
where the value of $\epsilon\,$ is taken from ~\eq{SolP1}. Thereby we see, 
that function ~\eq{Lyap1} is the Lyapunov function for the stable point
~\eq{StabP11}. A Lyapunov function for the stable point ~\eq{StabP2} could be constructed similarly.

\subsection{Lyapunov function for ~\eq{StabP3} stable point}

For the stable points ~\eq{StabP3}
\beq\label{StabP33}
\Le\,q_{St}\,,p_{St}\,\Ra_{3}\,=\,\Le\,\frac{1}{3\,\lambda}\,,\,\frac{1}{3\,\lambda}\,\Ra\,
\eeq
and corresponding asymptotic solution ~\eq{SolQ3} and ~\eq{SolP3} as a Lypunov we consider the following function
\beq\label{Lyap2}
F(q,p)\,=\,\ln\Le\,H\,/(\lambda\,q^{2}\,p)\Ra
\eeq
where $H$ is the Hamiltonian of \eq{En}. This function has following properties.
\begin{enumerate}
\item 
In the neighborhood of the stable point, where $q\,=\,\frac{1}{3\lambda}\,-\,\epsilon\,$
and $p\,=\,\frac{1}{3\lambda}\,-\,\epsilon\,$ with $\epsilon\,<\,1$, see \eq{SolQ3} and ~\eq{SolP3}, we have in leading order on $\epsilon$:
\beq\label{Prop1}
F(q,p)\,=\,\ln\Le\,1\,+\,9\,\epsilon\,\lambda\,\Ra\,\approx\,\,9\,\epsilon\,\lambda\,\,>\,0\,
\eeq
with 
\beq
\epsilon\,=\,\frac{e^{-Y/3}}{3\lambda}\frac{A\,-\,1/3\lambda}{A\,-\,2/3\lambda}\,.
\eeq
It is interesting to note also, that for
\beq\label{LCon}
1/3\,<\,A\lambda\,<\,2/3
\eeq
condition \eq{Prop1} for the function \eq{Lyap2} is not satisfied.
\item 
Taking precise value of the stable point  ~\eq{StabP3} we see that
\beq
F(q=\frac{1}{3\lambda},p=\frac{1}{3\lambda})\,=\,\ln\,1\,=\,0\,.
\eeq
\item
Taking the derivative of this function over rapidity we obtain:
\beq
\Le\,\frac{dF(q,p)}{dy}\,\Ra\,=\,\dot{H}\,-\,2\,\frac{\dot{q}}{q}\,-\,\frac{\dot{p}}{p}\,.
\eeq
At leading order on $\epsilon$ we have that $\dot{H}\,=\,0\,$ and we obtain
\beq
\Le\,\frac{dF(q,p)}{dy}\,\Ra\,=\,-\,2\,\Le\,1\,-\,\lambda\,q\,-\,2\,\lambda\,p\,\Ra  
\,-\,\Le\,-\,1\,+\,\lambda\,p\,+\,2\,\lambda\,q\,\Ra\,. 
\eeq
At leading order we, therefore, have
\beq
\Le\,\frac{dF(q,p)}{dy}\,\Ra\,=\,\Le\,-1\,+\,3\,\lambda\,q\,\Ra_{q\,=\,\frac{1}{3\lambda}\,-\,\epsilon\,}\,=\,
-\,3\,\epsilon\,<\,0\,.
\eeq
\end{enumerate}
Thereby we see, that our function ~\eq{Lyap2} is a Lyapunov function for the stable point ~\eq{StabP3}.

\section{Periodic limited cycles of solution}

In order to show the presence of periodic cycles in asymptotic solutions of our equations, we back to the 
original Hamiltonian ~\eq{InHam}
\beq\label{InHam1}
H=\mu\,q\,p\,-\,\gamma\,p^2\,q\,-\,\gamma\,p\,q^2\,
\eeq
and determines the vertices of equation correspondingly to the QCD Pomeron vertices :
\beq
\mu\,=\,\alpha_{s}\,,\,\,\,\,\gamma\,=\,\alpha_{s}^{2}\,,
\eeq
where $\alpha_{s}$ is QCD coupling constant, see \cite{0dimsol}.
Introducing new rescaled fields
\beq
p\,\rightarrow\,\frac{p}{\alpha_{s}}\,,\,\,\,\,q\,\rightarrow\,\frac{q}{\alpha_{s}}\,,
\eeq
we obtain rescaled Hamiltonian:
\beq\label{InHam11}
H\,=\,\frac{1}{\alpha_{s}}\,\Le\,q\,p\,-\,p^2\,q\,-\,p\,q^2\,\Ra.
\eeq
Now we follow the results of article \cite{NonLin}. Consider the system of three
variables $q,p$ and $z$:
$$
\dot{q}\,=\,q\,\Le\,1-\,q\,-\,\alpha\,p\,-\,\beta\,z\,\Ra\,
$$
$$
\dot{p}\,=\,p\,\Le\,1-\,p\,-\,\beta\,q\,-\,\alpha\,z\,\Ra\,
$$
\beq\label{TriS}
\dot{z}\,=\,z\,\Le\,1-\,z\,-\,\alpha\,q\,-\,\beta\,y\,\Ra\,
\eeq
with additional condition $\alpha\,+\beta\,=\,2$. The dynamics of following function
\beq
F\,=\,p\,+\,q\,+\,z\,
\eeq
could be described with the use of  equations ~\eq{TriS}: 
\beq
\frac{dF}{dy}\,=\,F\,\Le\,1\,-\,F\,\Ra\,.
\eeq
At $Y\,\rightarrow\,\infty\,$ the solution of this equation is simply
\beq\label{Restric}
F\,\rightarrow\,1\,,\,\,\,\,p\,+\,q\,+\,z\,\rightarrow\,1\,.
\eeq
Therefore, at asymptotically large rapidity $Y$ possible solutions of ~\eq{TriS} lie on the triangle ~\eq{Restric},
i.e. all orbits at large enough rapidity set up on this triangle. Further we project this motion
on the $q-p$ plane by use of projection
\beq\label{Projec}
z\,=\,1\,-\,q\,-\,p\,
\eeq 
which has the place at large rapidity in system ~\eq{TriS}. Rewriting the equations ~\eq{TriS}
we obtain for $q-p$ plane dynamics:
\beq
\dot{q}\,=\,\frac{\alpha\,-\,\beta}{2}\,q\,\Le\,1\,-\,q\,-\,2\,p\,\Ra\,
\eeq
\beq
\dot{p}\,=\,-\,\frac{\alpha\,-\,\beta}{2}\,p\,\Le\,1\,-\,p\,-\,2\,q\,\Ra\,.
\eeq
We see, that these equations could be considered as equations of motion for the following  Hamiltonian:
\beq\label{InHam2}
H\,=\,\frac{\alpha\,-\,\beta}{2}\,\Le\,q\,p\,-\,p^2\,q\,-\,p\,q^2\,\Ra.
\eeq
Comparing ~\eq{InHam2} and ~\eq{InHam1} we conclude, therefore,  that our Hamiltonian of interests could be considered as 
projection of dynamics of the system of three variables ~\eq{TriS} on the plane $q-p$ in the limit of large rapidity $Y$ with the
\beq
\alpha\,=\,1\,+\,\frac{1}{\alpha_{s}}\,
\eeq
\beq
\beta\,=\,1\,-\,\frac{1}{\alpha_{s}}\,\footnote{It is interesting to note that the ~\eq{TriS} system is drastically
simplified in the limit $\alpha_{s}\rightarrow\,\infty$}.
\eeq
We conclude, therefore, that at large rapidity limit the dynamics of the system of two variables ~\eq{Motion1} and system of three variables ~\eq{TriS}
is the same. The systems are dual in the limit of large $Y$.

In order to illustrate the periodic solutions in ~\eq{TriS}, and therefore periodic solutions in \eq{Motion1},
we introduce an additional function:
\beq
W\,=\,q\,p\,z\,.
\eeq
The dynamics of this function is described by following equation:
\beq
\frac{d}{dy}\ln(W)\,=\,3\,-\,3\,W\,=\,3\,\frac{d}{dy}\ln(F)\,,
\eeq
that gives after the integration:
\beq
W(y)\,=W_{0}\,\Le\,\frac{F(y)}{F(0)}\,\Ra^{3}\,\stackrel{y\rightarrow\,\infty}{\longrightarrow}\,W_{0}\,\Le\,\frac{1}{F(0)}\,\Ra^{3}\,=\,C_{0}\,.
\eeq
We see, that overall dynamics at large $Y$ is restricted by the following conditions
\beq\label{Con1}
\,p\,+\,q\,+\,z\,=\,1\,
\eeq
and
\beq\label{Con2}
\,p\,q\,z\,=\,C_{0}\,.
\eeq
Conditions imposed on the solutions of the ~\eq{TriS} at large rapidity result as cycle orbits, which are intersection of the triangle  ~\eq{Con1} and hyperboloid ~\eq{Con2}. In this case, the projection ~\eq{Projec}, determines also
cycle dynamics on the plane defined by equations of motion ~\eq{Motion1}.

The duality of the systems ~\eq{Motion1} and ~\eq{TriS} could be also understood from the point of conditions needed
to apply on the systems in order to obtain the solution. In the case of system ~\eq{Motion1} we need the boundary conditions
~\eq{InCon} and value of the energy of the system $E$. In the case of equation ~\eq{TriS} we need only boundary (initial) 
conditions $(q_{0}\,,\,p_{0}\,,\,z_{0}\,,\,)$. Excluding $z$ from ~\eq{Projec} and ~\eq{Con2} we obtain:
\beq
p\,q\,\Le\,1\,-\,q\,-\,p\,\Ra\,=\,C_{0}\,=\,E\,.
\eeq
We see that the following relation for the dual systems exists:
\beq\label{EnCon}
E\,=\,\,C_{0}\,=\,\frac{\,q_{0}\,p_{0}\,z_{0}\,}{\,q_{0}\,+\,p_{0}\,+\,z_{0}\,}\,.
\eeq
Hereby we conclude, that the energy parameter in system ~\eq{Motion1} is determined by the initial 
value $z_{0}$ in the system ~\eq{TriS} at given  $(q_{0}\,,\,p_{0}\,)$.

In our particular example for stable points ~\eq{Sol1} at $E\,=\,0$ we see, that
$\,p_{0}(y=0\,,\,Y\rightarrow\,\infty)\,\rightarrow\,0$  provides $E\,=\,C_{0}\,=\,0$ for corresponding asymptotic solution  
~\eq{FullSolQ1}- ~\eq{FullSolP1} through ~\eq{EnCon}.
In this case, in order to satisfy ~\eq{Con1}-~\eq{Con2} we also need to conclude that $\,z(\,Y\rightarrow\,\infty)\,\rightarrow\,0\,$.
In general, therefore, whole dynamics of the system ~\eq{Motion1} at large rapidity and at $E\,=\,0$ is described by the line
\beq
\,q(y)\,+\,p(y)\,=\,1\,
\eeq
with stable points given by ~\eq{StabP1}-~\eq{StabP2}. In the case when $E\,\neq\,0$ at the limit of large $Y$, the cycles are not lines anymore 
but they are cycles determined by ~\eq{Con1}-~\eq{Con2}.

\section{Conclusion}

In this paper we investigated the form and behavior of solutions of RFT-0 theory at the limit 
of asymptotically large rapidities. We developed a method which allows
to find an analytical form of asymptotic solution basing on the analysis of
steady states of the equations of motion of corresponding  Hamiltonian. Found solutions
satisfy the boundary conditions of the problem, \eq{InCon}, and preserve the symmetry of the problem, 
\eq{SymCon}, and therefore, they could be considered also as the approximate solutions of our equations of 
motion in the limit of large rapidity. The results obtained are important because solutions of RFT based on 
QCD Pomerons exhibits similar properties 
and have similar structure due the similarities of equations of motion, see \cite{BonMot}.

The pairs of the functions \eq{FullSolQ1}-\eq{FullSolP1} and \eq{FullSolQ2}-\eq{FullSolP2}
were known a long time ago, see \cite{Fan}, as "fan" diagrams solutions. They are good approximation
for the scattering amplitude in the case when the scattering situation is not symmetrical, see
\cite{Braun,BonMot}. Therefore, it was interesting to check, whether the same functions provide minimal energy
for the Hamiltonian also in the case of symmetrical scattering with conditions \eq{InCon}, see "fan"
dominance effect discussion in \cite{BonMot}. Our results, thereby, prove that in the semi-classical
approximation the "fan" diagrams are the leading contribution into the scattering amplitude at the limit of large rapidity also in the case of scattering of symmetrical objects. The stability of these solutions was  
shown as well by use of the usual stability analysis, namely by the Lyapunov functions construction, see \eq{Lyap1}.

Considering the symmetrical solution \eq{SolQ3}-\eq{SolP3} we see that this solution is suppressed
in the amplitude, see \cite{BonMot}, in comparison with the non-symmetrical, providing non-zero Hamiltonian.
Therefore, in calculations of different observables of RFT based on QCD Pomerons, \cite{Braun,BonMot}, in the
first approximation the leading contribution comes from the "fan" diagrams. The symmetrical solution
could be considered, therefore, as sub-correction to the amplitude on the semi-classical level. 
The stability of this solution is checked by the construction of
Lyapunov function \eq{Lyap2}, which was obtained in assumption that condition \eq{LCon} is not satisfied. Is this
a sign of instability of this solution for some values of parameters is not clear. 
 
Another result of our investigation is a presence of limit cycles solutions which could
be demonstrated on the base of the system of equations dual to the considered one in the limit of large rapidity.
In order to understand possible consequences of such solutions on observables of scattering processes in RFT 
at high energies an additional investigation is required. Also, in the future studies, it will be important to
understand a mechanism of reduction of the limit cycle to the one from the stable points. 

A problem of an origin
of three solutions during the evolution of the system with rapidity is a subject which was not fully
considered in the paper. A simplest analysis shows, that
new solution are arose when the the function $p$ or $q$ begins to be large enough. It happens approximately 
at rapidity $y\propto\,-\,\frac{1}{\mu}\ln(\lambda\,A)\,\rightarrow\,1$, but we have no dynamical
description of the process similar to usual bifurcation picture. 
It have be noted also, that more complicated boundary conditions lead to more complicated picture
of possible solutions of the equations of motion, see \cite{BBraun}. A rise of these solutions it is 
an interesting problem which we plan investigate in our further work as well, as the problems mentioned above.

\end{document}